\documentclass[10pt,a4paper,abstracton]{scrartcl}
\usepackage{cite}
\usepackage{amsmath,amssymb,graphicx,authblk,multicol}

\usepackage{geometry}

\title{Cooling of Compact Stars with Color Superconducting Quark Matter\thanks{Submitted to ACTA ASTRONOMICA SINICA, Proceedings of Quarks and Compact Stars (QCS 2014)}}

\author[1]{Tsuneo~Noda\thanks{noda@kurume-it.ac.jp}}
\author[2]{Nobutoshi~Yasutake}
\author[3]{Masa-aki~Hashimoto}
\author[4]{Toshiki~Maruyama}
\author[5]{Toshitaka~Tatsumi}
\author[6]{Masayuki~Y.~Fujimoto}

\affil[1]{Kurume Institute of Technology, Fukuoka, 830-0052 JAPAN}
\affil[2]{Physics Department, Chiba Institute of Technology, Chiba, 275-0023 JAPAN}
\affil[3]{Department of Physics, Kyushu University, Fukuoka, 812-8581 JAPAN}
\affil[4]{Advanced Science Research Center, Japan Atomic Energy Agency, Ibaraki, 319-1195 JAPAN}
\affil[5]{Department of Physics, Kyoto University, Kyoto, 606-8502 JAPAN}
\affil[6]{Department of Physics, Hokkaido University, Sapporo, 060-0810 JAPAN}

\date{}

\begin{document}

\maketitle

\begin{abstract}
We show a scenario for the cooling of compact stars considering the central source of Cassiopeia A (Cas A).
The Cas A observation shows that the central source is a compact star with high effective temperature, and it is consistent with the cooling without exotic phases.
The Cas A observation also gives the mass range of $M \geq 1.5 M_\odot$.
It may conflict with the current cooling scenarios of compact stars that heavy stars show rapid cooling.
We include the effect of the color superconducting  (CSC) quark matter phase on the thermal evolution of compact stars.
We assume the gap energy of CSC quark phase is large ($\Delta \gtrsim \mathrm{10 MeV}$),
and we simulate the cooling of compact stars.
We present cooling curves obtained from the evolutionary calculations of compact stars: while heavier stars cool slowly,
and lighter ones indicate the opposite tendency.
\end{abstract}

{\bf Keywords:} dense matter --- stars: neutron

\begin{multicols}{2}
\section{Introduction}

Cooling theory of compact stars has been discussed for decades.
It has been believed that some compact stars require exotic cooling to explain the observations and others can be explained by Modified URCA and Bremsstrahlung processes (the standard cooling).
The exotic state appears at high density region, and it can be larger than the central density of light compact stars.
Therefore, only heavy stars have the exotic phase in their cores and they cool faster than lighter stars.

The scenario seemed to be fine until the observation of the central source of Cas A\cite{hh09}.
The observation of Cas A shows that it is a heavy compact star and its surface temperature is relatively high.
There were no observation of isolated compact star that has the mass and the temperature together.
It satisfies the standard cooling scenario, it seems that all exotic scenarios ruled out.
However, some compact stars (e.g., SAX J1808, 3C38 and Vela pulsar) are difficult to be explained by the standard scenarios.
It is possible to think these stars have much heavier than Cas A, but it may conflict with current supernovae theory.

Recently, the surface temperature of Cas A becomes an issue.
Ref. \cite{hh10} reported the surface temperature of Cas A is decreasing in the past 10 years.
The temperature drop is too rapid to explain by usual neutrino emission process, except superfluidity of neutrons.
Recent studies (such as \cite{5,7,TN13}) insist that the rapid decrease in the surface temperature shows that the transition to neutron superfluidity occurs.
However, the observation has been re-analyzed\cite{el13, pp13}, and the temperature drop is still under discussion.

In this study, we present a model which satisfies the temperature-mass relation of Cas A and the temperatures of other stars,
by considering color superconducting (CSC) quark phase in the core of neutron stars.

\section{Cooling Models}

We construct a model that includes both quark-hadron mixed phase (MP) and CSC phase.
We use the EoS of the Bruekner-Hartree-Fock theory (hadron) and the Dyson-Schwinger theory (quark), including the MP phase between the both phases\cite{NY}. We assume that the entire quark matter (QM) is in the CSC phase with large energy gap ($\Delta \gtrsim 10~\mathrm{MeV}$). The maximum mass with this EoS is $2.13 M_\odot$, and it satisfies the recent $2 M_\odot$ observations\cite{de10, an13}.

The EoS of the hadronic phase is stiff enough to obtain large proton fraction $y_p$. It becomes larger than the threshold proton fraction of Direct URCA process ($y_p > 1/9$), therefore we need to consider the strong Direct URCA process.
To suppress this strong process, we assume that strong proton superfluidity ($T^\mathrm{p}_{\mathrm{CR}} \gtrsim 2 \times 10^9~\mathrm{K}$) works.

Another important phase in compact stars is the neutron $^3P_2$ superfluidity.
The critical temperature of the triplet phase $T^{\mathrm{n3}}_{\mathrm{CR}}$ is comparable with the matter temperature during the neutrino cooling phase.
Therefore the transition to superfluid phase may occur in this duration.
The neutron superfluidity has two effects, one is to suppress the neutrino emissivity in the superfluid state, and the other is to emit large number of neutrinos at the transition to the superfluid state. The emission effect, known as ``PBF'' (Pair Breaking and Formation), causes rapid drop of the stellar temperature, and it is sensitive to the critical temperature.
For simplicity and the uncertainty of Cas A observation, we do not include the effect of neutron superfluidity for our calculation.

\end{multicols}
\begin{figure}[tbph]
	\begin{center}
		\includegraphics[width=0.5\linewidth,keepaspectratio,clip]{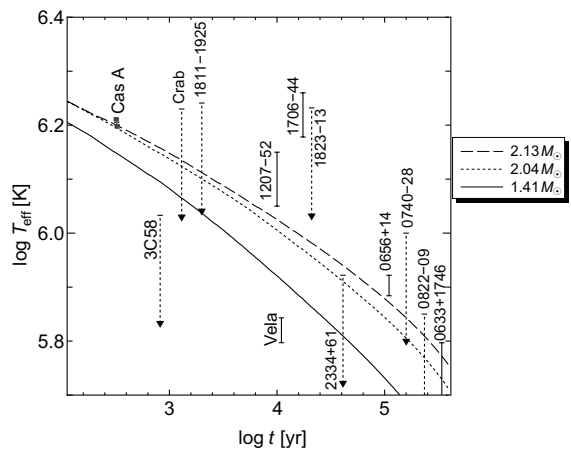}
	\end{center}
	\caption{Cooling curves with CSC quark phases. Vertical lines indicate the observational data.}
	\label{fig:CC}
\end{figure}
\begin{multicols}{2}

We select stellar masses of compact stars to be $1.41$, $2.04$, and $2.13 M_\odot$.
The results are shown in Fig. \ref{fig:CC} with available observational vales (see Ref. \cite{TN13}).
We can see the cooling curves transit from cooler regions to the hotter regions with increasing the mass of compact star.
It can be said that the compact star with larger mass cools slower than lighter star.
The tendency is opposite to the current cooling scenarios.
The cooling curves do not seem to satisfy the data of 3C58 or Vela pulsar. To explain these observational data, the neutron $^3P_2$ superfluidity is required for further calculation.

\section{Discussions}

We demonstrate the effect of CSC quark phase on the cooling curve with the EoS which satisfies observations of the $2 M_\odot$ compact stars.
It is possible to make a situation that heavy star cools slower than lighter stars.

To understand the thermal evolution of compact stars, considering 3 super phases (neutron  $^3P_2$, proton  $^1S_0$ superfluidity and quark CSC) is important.
Quark phase in compact stars can be in CSC phase. Once the CSC phase appears, it suppress the strong neutrino emission by quarks.
In the hadronic phase, neutron and proton superfluidity occur at particular density regions.
The proton superfluidity suppresses neutrino emission. The neutron superfluidity plays the both roles, rapid cooling during transition, and suppression to the neutrino emissivity after the transition.
The theories of these super phases are still uncertain. Cooling calculation of compact stars can connect nuclear theories of high density region and observations.

\end{multicols}
\end{document}